 \documentclass[preprint2]{aastex}
 \usepackage{natbib,aas_macros}
 \usepackage[usenames,dvipsnames]{color}
 \usepackage{amsmath}
\shorttitle{Sensitivity of helioseismic travel-times to the imposition of a Lorentz force limiter in computational helioseismology}
\shortauthors{Moradi \& Cally}

\begin{document}


\title{Sensitivity of helioseismic travel-times to the imposition of a Lorentz force limiter in computational helioseismology}

\author{Hamed Moradi and Paul S. Cally}
\affil{Monash Centre for Astrophysics, School of Mathematical Sciences, Monash University, Clayton, \\Victoria 3800, Australia}

\email{hamed.moradi@monash.edu}

\begin{abstract}
The rapid exponential increase in the Alfv\'en wave speed with height above the solar surface presents a serious challenge to physical modelling of the effects of magnetic fields on solar oscillations, as it introduces a significant CFL time-step constraint for explicit numerical codes. A common approach adopted in computational helioseismology, where long simulations in excess of 10 hours (hundreds of wave periods) are often required, is to cap the Alfv\'en wave speed by artificially modifying the momentum equation when the ratio between Lorentz and hydrodynamic forces becomes too large. However, recent studies have demonstrated that the Alfv\'en wave speed plays a critical role in the MHD mode conversion process, particularly in determining the reflection height of the upward propagating helioseismic fast wave. Using numerical simulations of helioseismic wave propagation in constant inclined (relative to the vertical) magnetic fields we demonstrate that the imposition of such artificial limiters significantly affects time-distance travel times unless the Alfv\'en wave-speed cap is chosen comfortably in excess of the horizontal phase speeds under investigation.

\end{abstract}


\section{Introduction}\label{intro}

Sunspots are large cool regions on the solar surface associated with strong magnetic flux concentrations. They represent an important connection of the internal magnetic field of the Sun with the solar atmosphere. Their subsurface structure is a challenge to both theory and observation, and in particular to local helioseismology, which cannot yet fully deal with the dominant magnetic fields near the surface \citep[for recent reviews see][]{moradi2012,gbs2010,moradietal2010}. The ultimate aim is to better understand solar activity by seismically probing flows, magnetic fields, and thermal structures in sunspots and the wider active regions that host them.

A major issue in local helioseismology is the complex behaviour of trapped acoustic waves ($p$-modes) when they encounter an active region \citep{cb1993,cbz1994,sc2006,cally2007,cg2008,schunkeretal2013}. The magnetic field turns them into a complex mixture of fast, slow and Alfv\'en waves through MHD mode conversion processes. The trapped modes can then partially escape into the solar atmosphere and partially reflect there to rejoin the internal wave field, with significant consequences for the local seismology. 

Physical modelling of the effects of magnetic fields on solar oscillations can help explain helioseismic measurements in regions of strong magnetic field. Through forward modelling -- the art of constructing computational models that mimic wave propagation through sunspots and matching the resultant wave statistics with observations -- we have been able to gain much valuable insight into the interaction of helioseismic waves with magnetic fields \citep[e.g., ][]{cgd2008,cameronetal2011,hanasoge2008,khomenkoetal2009,pk2009,shelyagetal2009,mhc2009,schunkeretal2013}. 

Simulating 3-D MHD wave propagation in the Sun is a complex and time-consuming task however, with one of the most pressing issues being how to treat the excessively large Alfv\'en wave speed ($c_a = B_0/\sqrt{4\pi\rho_0}$; where $B_0$ and $\rho_0$ represent the magnetic field strength and density respectively) above the surface, which is brought about by the exponential drop in $\rho_0$ with height. With $c_a$ reaching several thousands of $\rm km\,s^{-1}$, this leads to an extremely stiff numerical problem for the explicit numerical solvers, as the time-step ($\Delta t \approx \Delta z/c_a$, where $\Delta z$ denotes the vertical grid resolution) is constrained by the Courant-Friedrichs-Lewy (CFL) condition, resulting in the need for very small $\Delta t$ for simulating even moderate magnetic field strengths. 

Consequently, in order to simulate helioseismic data sets in a feasible amount of time, the most common approach in computational helioseismology has been to introduce a Lorentz force ($\mathbf{F}_L$) scaling factor that limits it when the ratio between Lorentz and hydrodynamic forces (or in other words $c_a/c_s$, where $c_s$ is the sound speed) becomes too large. Typically $\mathbf{F}_L$ is scaled by $\alpha c_s^2/(\alpha c_s^2 + c_a^2)$ (where $\alpha$ is a free parameter that controls the amplitude of the limiter), resulting in $c_a$ being capped above the surface, commonly in the range of $\sim 20-60$ $\rm km\,s^{-1}$ \citep{rsk2009,cameronetal2011,braunetal2012}. Another similar approach has been to scale the magnetic field by a pre-factor such that $c_a$ never exceeds a certain predefined value \citep{hanasogeetal2012}. The physical implications of artificially limiting $c_a$ in such manners have not been fully explored to date, with the general assumption being that the overlying atmosphere does not play a significant role in the seismology of the subphotosphere. 

However, a number of recent studies have cast doubt on this assumption by demonstrating the critical role played by the exponentially increasing $c_a$ in the fast-to-Alfv\'en mode conversion process which takes place in the lower solar atmosphere \citep{cg2008,ch2011,hc2012, kc2011, kc2012, felipe2012}. Using idealised MHD simulations, these studies show how the upward propagating helioseismic (fast) wave can reflect off the $c_a$ gradient back to the surface at the `fast-wave reflection height' where the horizontal phase speed ($v_{ph} = \omega/k_h$; where $\omega$ denotes angular frequency and $k_h$ the horizontal wave number) roughly coincides with $c_a$. These 3-D calculations also show that, depending on the local relative inclinations and orientations of the background magnetic field and the wavevector, the fast wave may undergo partial mode-conversion to either an upward or downward propagating Alfv\'en wave \citep[see Figure 1 in][]{kc2012} around the reflection height where they are near-resonant. 

While the fast-to-Alfv\'en mode conversion process generally takes place above the surface, it has potentially serious consequences for helioseismology. This is because after they reflect off the $c_a$ gradient, the fast waves re-enter the solar interior wave field, meaning that their journey through the atmosphere and the phase changes they suffer in the conversion process must have some effect on the seismology. Inversions of observed time-distance travel times \citep[e.g.,][]{kds2000,couvidatetal2005} would normally but mistakenly interpret such phase changes as `travel time shifts' due to subsurface inhomogeneities.

In a recent follow-up study, \citet{cm2013} quantified the implications of the returning fast and Alfv\'en waves for the seismology of the photosphere. They found substantial wave travel time shifts that acutely depend on magnetic field inclination and wave propagation orientation, in direct correspondence with the escaping acoustic and Alfv\'enic wave fluxes above the surface. However in another related study, it was observed that the imposition of a $\mathbf{F}_L$ limiter actually suppresses the fast-to-Alfv\'en mode conversion process by artificially diminishing the Lorentz force above the surface \citep{mc2013}. But nonetheless, reality (and indeed, finite computational resources) dictates that simulating artificial data sets on the time scales required for local helioseismic analysis will ultimately necessitate the use of a limiter to ensure a manageable CFL condition. Whether this has a flow-on effect to the helioseismic travel times is something that we wish to explore here, by modelling wave propagation through constant inclined magnetic fields and comparing the artificial helioseismic travel times derived from simulations with and without a $\mathbf{F}_L$ limiter. 

\section{Numerical Model}\label{modelling}

We use the Seismic Propagation through Active Regions and Convection \citep[SPARC;][]{hanasoge2007} code to conduct the forward modelling component of our analysis. SPARC solves the 3-D linearised MHD equations in Cartesian geometry to investigate wave interactions with local perturbations. We employ a similar numerical setup to \citet{cm2013}, with a 3-D computational box spanning 26.53 Mm in height $z$ (covering $-25 \leqslant z \leqslant 1.53 $ Mm) using $265$ vertical grid points, with the grid spacing ranging from several hundred km deep in the interior to tens of kms in the near-surface layers. The box length is chosen to be 140 Mm in the horizontal directions $x$ and $y$. We use $128$ evenly spaced grid points in $x$ and $y$, resulting in a horizontal resolution of $\Delta x = \Delta y= 1.09$ Mm/pixel. The vertical and horizontal boundaries of the box are absorbent, with perfectly matched layer (PML) boundary layers at the top and bottom and absorbing sponges lining the sides.
\begin{figure}[ht!]
\centering
\includegraphics[width=0.45\textwidth,trim=1cm 1cm 1cm 1cm, clip=true]{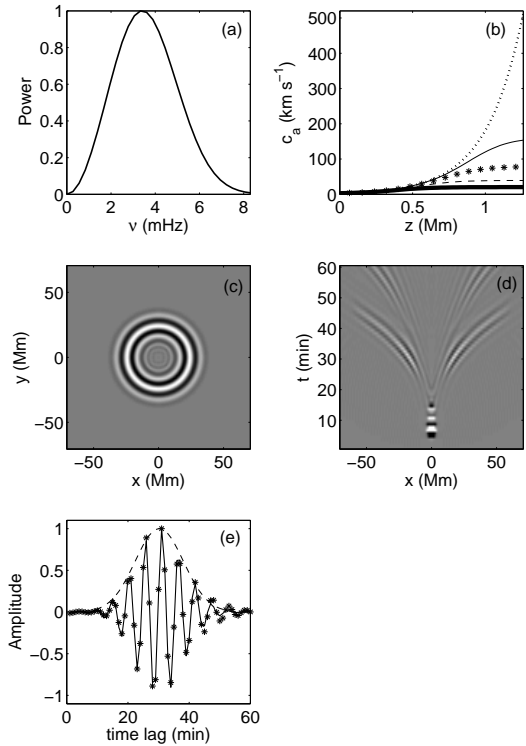}
\caption{(a) Represents the normalised power (arbitrary units) vs.~frequency spectrum of the acoustic wave source. (b) Shows $c_a$ as a function of height above the photosphere ($z=0$) in the absence of a $\mathbf{F}_L$ limiter (dotted line), and with a limiter imposed to cap $c_a$ at 20 km s$^{-1}$ (bold solid line), 40 km s$^{-1}$ (dashed line), 80 km s$^{-1}$ (asterisks), and finally at 160 km s$^{-1}$ (light solid line). (c) Shows a snapshot of the $v_z$ wave field in the quiet-Sun simulation at t = 35 minutes. (d) Represents the time-distance diagram produced by taking a cut at $y=0$ in the quiet-Sun simulation. (e) Shows an example of a Gabor wavelet fit (solid line) to the wave form (asterisks) produced by the $\theta=50\degr$ simulation (without a limiter imposed), for $\Delta = 11.6$ Mm and $\phi = 0\degr$. The dashed lines gives the envelope of the fitted wavelet. }
\label{fig_4plots}
\end{figure}

The background model consists of a convectively stabilised solar model \citep[CSM\_B;][]{schunkeretal2011} threaded by uniform magnetic field of $B_0=500$ G, inclined at angle $\theta$ (inclination from vertical, with $0\degr \leqslant \theta \leqslant 90\degr$). In \citet{cm2013} we employed random stochastic wave sources to generate acoustic waves, but in this study we employ a perturbation source in vertical velocity ($v_z$) similar to \citet{shelyagetal2009}: 
\begin{equation}
v_z = \sin \frac{2\pi t}{t_1} \exp \left( -\frac{(r-r_0)^2}{\sigma_r^2} \right) \exp \left( - \frac{(t-t_0)^2} {\sigma_t^2} \right), 
\end{equation}  
where $t_0 = 300 $s, $t_1= 300 $s, $\sigma_t = 100 $s, $\sigma_r = 4\Delta x$, and $r_0$ is the source position, located 160 km below the surface at $(x_0 ,y_0) = (0, 0)$.  
This source generates a broad spectrum of acoustic waves in the $3.33$ mHz range (see Figure \ref{fig_4plots} a), mimicking wave excitation in the Sun.  

We initially use the acoustic source to simulate wave propagation for ten different field inclinations ($\theta = 0\degr, 10\degr, 20\degr..., 90\degr$), without invoking a limiter. The time-step required for each of these simulations in SPARC is $\Delta t = 0.1$ s. We then repeat the simulations using the form of the limiter adopted by \citet[][]{hanasogeetal2012} to cap $c_a$ at a number of values above the surface. The momentum equation which results is thus:
\begin{eqnarray}
\partial_t \mathbf{v} =-\frac{1}{\rho_0}\nabla p - \frac{\rho}{\rho_0}g\,\mathbf{\hat{e}}_z  + \nonumber \\  \frac{ [\left (\nabla \times \mathbf{B}^*_0) \times \mathbf{B} + (\nabla \times \mathbf{B}) \times \mathbf{B}^*_0 \right]}{4\pi\rho_0} + \mathbf {S},
\end{eqnarray}
where $\rho_0$ denotes density (the subscript ÔÔ0ÕÕ indicates a time-stationary background quantity, whereas unsubscripted terms fluctuate), $p$ the pressure, $\mathbf{B} ~$magnetic field, $\mathbf{B}^*_0 = \sqrt\kappa \mathbf{B}_0$ (where $\kappa = \alpha c_s^2/(\alpha c_s^2 + c_a^2)$), $g$ gravity (with direction $-\mathbf{\hat{e}}_z$), and $\mathbf{S}$ the source term.

The first cap is placed at 20 km s$^{-1}$ ($\alpha=8$, resulting in $\Delta t = 2.0$ s), followed by 40 km s$^{-1}$ ($\alpha = 30$, $\Delta t = 1.0$ s), 80 km s$^{-1}$ ($\alpha = 125$, $\Delta t =0.5$ s), and finally 160 km s$^{-1}$ ($\alpha = 540$, $\Delta t = 0.25$ s). Figure \ref{fig_4plots} b) plots the various $c_a$ profiles of each simulation. Hence, in total, we conduct 50 unique simulations, each with a temporal duration of 1 hour. In addition to the magnetic simulations, we also complete a quiet-Sun reference simulation, using only the vertical stratification of the thermodynamic parameters in the CSM\_B model. 

For the proceeding helioseismic analysis we use one hour simulated $v_z$ data cubes extracted at a constant geometrical height of 300 km above the surface.

\section{Time-Distance Analysis}\label{td}

In time-distance helioseismology wave travel times are generally calculated from fits to the temporal cross correlation function between two points (source and receiver) at the solar surface \citep{duvalletal1993}. However since wave excitation in our computational box is generated by a single source, there is no need to compute the velocity correlations between different surface points, meaning that time-distance diagrams can be constructed by plotting $v_z$ as functions of time for all horizontal points (see Figures \ref{fig_4plots} c, d). Moreover, since the simulations are in 3-D, by simply selecting a receiver point away from the central axis in the $xy$-plane, we have the ability to choose the magnetic field orientation with respect to the vertical plane of wave propagation, which we refer to as the ``azimuthal'' field angle ($\phi$, where $0\degr \leqslant \phi \leqslant 180\degr$)\footnote{In \citet{cm2013}, since random wave excitation was employed, Fourier filtering was applied in wavevector space to isolate particular azimuthal directions.}.   

Prior to calculating the travel times, we first filter the data cubes in two frequency ranges: 3 and 5 mHz by employing a Gaussian filter with a dispersion of 0.5 mHz. We then measure the phase travel time perturbations $\delta \tau$ (i.e., the differences in the phase travel times between the magnetic and nonmagnetic simulations) using Gabor wavelet fits \citep{kd1997} to the time-distance diagram at particular wave travel distances ($\Delta$) away from the source, as functions of field inclination ($\theta$) and azimuthal direction ($\phi$). A rectangular window of width 14 minutes centred on the first-bounce ridge selects the fitting interval in time lag. The fits are done by minimising the misfit between the Gabor wavelet and the wave form. Figure \ref{fig_4plots} e) shows an example of a wave form and the associated best-fit Gabor wavelet from one of our calculations.

We measured $\delta \tau$ for a range of distances, but for the sake of brevity we present the results for $\Delta=11.6$, $24.35$ and $42.95$ Mm below. The horizontal phase speeds $v_{ph}$ associated with these distances are 16.3, 34.8 and 46.8 km s$^{-1}$ respectively\footnote{For $p$-modes in the ray approximation, $v_{ph}$ is equal to the sound speed at the lower turning point of a ray that travels a horizontal distance $\Delta$ (neglecting the magnetic field and acoustic cutoff effects).}. 

\section{Results \& Analyses}

\begin{figure}[ht!]
\centering
\includegraphics[width=0.45\textwidth,trim=1cm 1cm 1cm 0cm, clip=true]{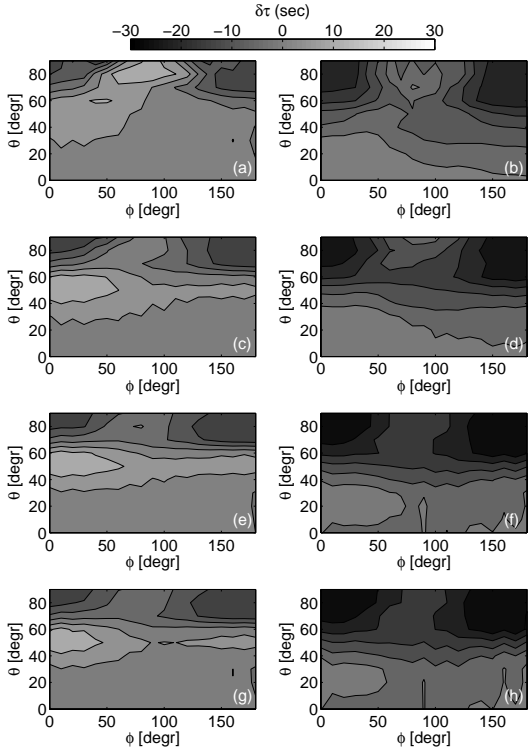}
\caption{Contour plots of $\delta\tau$ (in seconds) as functions of $\theta$ and $\phi$ (in degrees) for $\Delta = 11.6$ Mm. Left hand column represents 3 mHz and right 5 mHz. Panels (a-f) are derived from simulations using a limiter with a $c_a$ cap at 20 (a, b), 40  (c, d), and 160 (e, f) km s$^{-1}$, while panels (g, h) are derived without a limiter. The oscillations seen in the contour lines are artefacts of the coarse resolution and noise in the time-distance fittings.}
\label{fig_r11}
\end{figure}

\begin{figure}[ht!]
\centering
\includegraphics[width=0.45\textwidth,trim=1cm 1cm 1cm 0cm, clip=true]{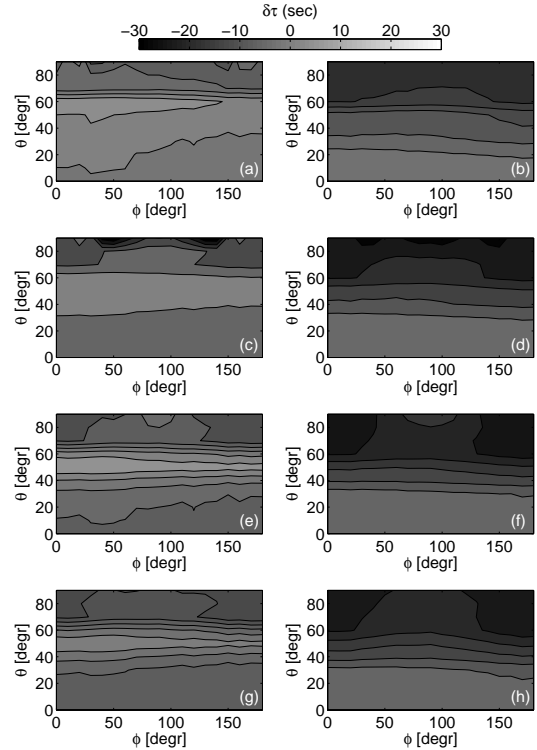}
\caption{Same as Figure \ref{fig_r11} but for $\Delta = 42.95$ Mm.}
\label{fig_r39}
\end{figure}

The contour plots in Figures \ref{fig_r11} and \ref{fig_r39} show the resulting 3 (left column) and 5 (right column) mHz $\delta\tau$ for $\Delta = 11.6$ and $42.95$ Mm respectively, as functions of $\theta$ and $\phi$. The bottom panels (g, h) of each figure represent the results derived from the simulations in which we did not impose a  limiter. The general pattern of $\delta\tau$ for these cases are strikingly similar to those presented in \citet{cm2013}, with the behaviour being strongly linked to mode conversion in the atmosphere (see their Figures 2 and 5). A detailed discussion is presented in \citet{cm2013}, which we summarise here in short: 
\begin{itemize}
\item At low field inclinations \citep[insufficient to provoke the ramp effect, see e.g., ][]{bl1977,jefferiesetal2006} the upward propagation of acoustic waves into the atmosphere is severely inhibited due the acoustic cutoff frequency ($\omega_c$, being just over 5 mHz in the atmosphere), resulting in $\delta \tau$ values of a few seconds being recorded. 
\item However, once $\omega > \omega_c \cos \theta$, the atmosphere is open to acoustic wave penetration, which results in substantial negative $\delta\tau$ at small $\sin \phi$ (i.e, for $\phi \lesssim 30\degr$ and $\gtrsim 150\degr$).  
\item At intermediate $\phi$, the fast wave loses more energy near its apex to the Alfv\'en wave, contributing a positive $\delta\tau$ that partially cancels the underlying negative $\delta\tau$. 
\end{itemize}

\begin{figure}[ht!]
\centering
\includegraphics[width=0.45\textwidth,trim=1cm 1cm 1cm 1cm, clip=true]{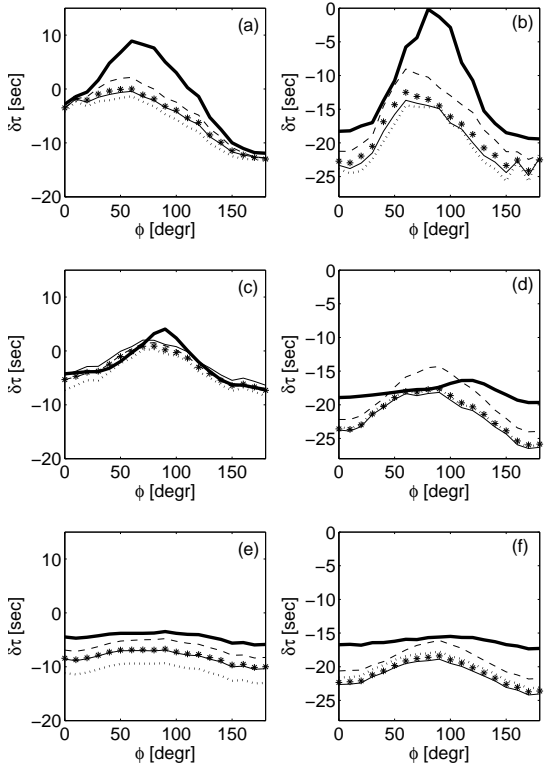}
\caption{ $\delta\tau$ (in seconds) for $\theta = 70\degr$ plotted as functions of $\phi$ (in degrees) derived from simulations without a $\mathbf{F}_L$ limiter (dotted line), and with limiters imposed capping $c_a$ at 20 (bold solid line), 40 (dashed line), 80 (asterisks) and 160 (light solid line) $\rm km\,s^{-1}$. Panels (a, b) represent $\Delta = 11.6$ Mm, (c, d) $\Delta = 24.35$ Mm and (e, f) $\Delta = 42.95$ Mm. Left column represents 3 mHz and right 5 mHz. Note that $5.2\cos70\degr=1.8$, so even 3 mHz is well above the ramp-reduced atmospheric cutoff at these high inclinations.}
\label{fig_td70}
\end{figure}

\begin{figure}[ht!]
\centering
\includegraphics[width=0.45\textwidth,trim=1cm 1cm 1cm 1cm, clip=true]{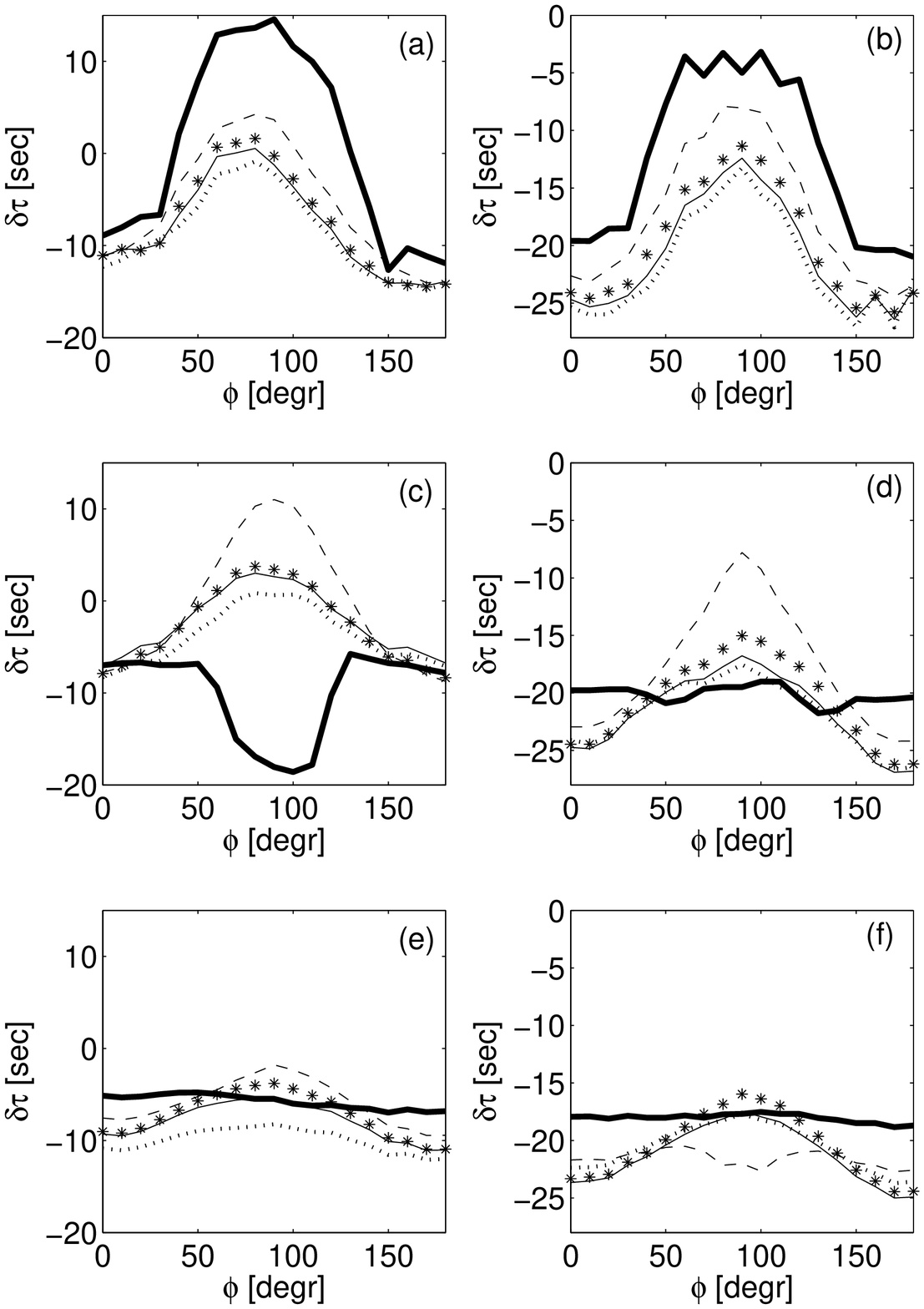}
\caption{Same as Figure \ref{fig_td70} but for $\theta = 80\degr$. }
\label{fig_td80}
\end{figure}

Panels (a-f) in Figures \ref{fig_r11} and \ref{fig_r39} show the $\delta\tau$ derived from simulations which have $\mathbf{F}_L$ limiters imposed to cap $c_a$ at 20 (a, b), 40 (c, d) and 160 (e, f) km s$^{-1}$. When comparing the overall $\delta\tau$ behaviour in these panels to those derived in the absence of a limiter (g, h), we notice distinctive differences in both the magnitude and general behaviour of $\delta\tau$ across both $\theta$ and $\phi$, evident at both 3 and 5 mHz and being most prevalent around $\theta \approx 60\degr-90\degr$ and $\phi \approx 90\degr$ (a similar $\delta\tau$ pattern was also observed for $\Delta = 24.35$ Mm, but is not shown here). This can be seen more clearly in Figures \ref{fig_td70} and \ref{fig_td80}, which plot $\delta\tau$ for $\theta = 70\degr$ and $80\degr$ across $\phi$, for all three $\Delta$ and $\mathbf{F}_L$ limiters studied. Differences of up to 20 seconds in $\delta\tau$ can be observed around $\phi=80\degr$ with the $c_a$ cap at 20 $\rm km\,s^{-1}$ . The situation improves somewhat as the cap is lifted to 40 km s$^{-1}$, but differences of up to $\sim 10$ seconds still persist for some distances. But as the $c_a$ cap is raised to 80 and then 160 km s$^{-1}$, the differences in $\delta\tau$ become progressively smaller and we observe a convergence to the $\delta \tau$ values derived without a limiter. We did not observe complete convergence, even with the $c_a$ cap at 160 $\rm km\,s^{-1}$. 

This behaviour is not completely unexpected. As discussed earlier, the fast wave reflection height is determined by where $v_{ph} \approx c_a$. This implies that by artificially limiting $c_a$ above the surface, we are allowing fast waves with $v_{ph}$ above $c_a$ to reach the absorbing PML layer at the top of the computational domain and therefore never to return to the subsurface seismic field. As is evident in Figures \ref{fig_r11}--\ref{fig_td80}, this has profound consequences for travel-time measurements if $c_a$ is capped below the $v_{ph}$ associated with the chosen $\Delta$. This explains why we observe the most significant differences in $\delta\tau$ associated with the $c_a$ cap at 20--40 km s$^{-1}$ (remembering that the largest $v_{ph}$ sampled, associated with $\Delta=42.95$ Mm, is 46.8 km s$^{-1}$), and at the relative inclinations and azimuthal orientations typically associated with maximal fast-to-Alfv\'en mode conversion \citep{cg2008,kc2011,kc2012,felipe2012}. Only once the $c_a$ is placed comfortably above the $v_{ph}$ being studied do we begin to see a convergence to the $\delta \tau$ derived without a limiter. Furthermore, since fast-to-Alfv\'en conversion is also spread over many scale heights for wavenumbers typical of local helioseismology \citep{ch2011}, it explains why we don't observe a complete convergence in $\delta\tau$, even with the $c_a$ cap at 160 km s$^{-1}$. 

It is also worth noting that even in some of the cases with the limiter set at 160 $\rm km\, s^{-1}$, the resulting $\delta\tau$ discrepancies of $\sim1-2$ seconds are comparable with the estimates from \cite{schunkeretal2013} for the sensitivity of travel times to changes in the subsurface structure of sunspots. In fact, even 160 $\rm km\,s^{-1}$ could be insufficient in some cases for practical helioseismology with large $\Delta$. Finally, we note that we also conducted a number of test cases using the same simulation setup as described in Section \ref{modelling}, but with $B_0 = 1$ and $1.5$ kG, and apart from larger amplitudes in $\delta\tau$, the overall results were almost identical to those derived from the $500$ G cases. 

\section{Summary \& Discussion}

Computational helioseismology generally entails high resolution, long (temporal) duration 3-D MHD simulations to study the interaction of helioseismic waves with magnetic fields. To alleviate the severe CFL time-step constraints introduced by the exceedingly high $c_a$ above the surface in magnetic regions, a number of explicit numerical codes limit the strength of $\mathbf{F}_L$ in low plasma-$\beta$ regions, essentially capping the $c_a$ gradient above the surface in the process. While this approach can increase the explicit time-step limit to any desired or practical value, it also severely impacts on the fast-wave reflection height ($c_a \approx v_{ph}$), and thus by extension the fast-to-Alfv\'en mode conversion process, which recent studies have shown to be problematic for helioseismology. 

Using 3-D MHD simulations of waves in homogenous inclined magnetic fields we find that, in the absence of a $\mathbf{F}_L$ limiter, time-distance $\delta\tau$ are sensitive to magnetic field inclination and wave propagation orientation, consistent with the recent results of \citet{cm2013}. We also find that the imposition of an artificial $\mathbf{F}_L$ limiter can have a significant impact on these $\delta\tau$, unless the $c_a$ cap is placed well above the horizontal phase speed associated with the wave travel distance being studied -- thus assuring that the reflection height of helioseismic fast wave in the lower solar atmosphere remains (relatively) unharmed. 

\acknowledgments

This research was undertaken with the assistance of resources provided by the NCI National Facility at the Australian National University, the Multi-modal Australian ScienceS Imaging and Visualisation Environment (MASSIVE), and the gSTAR national facility at Swinburne University of Technology. gSTAR is funded by Swinburne and the Australian Government's Education Investment Fund.


\clearpage

\end{document}